**Liquid-like Free Carrier Solvation and Band Edge Luminescence in Lead-Halide Perovskites**

Yinsheng Guo,[1] Omer Yaffe,[2] Trevor D. Hull,[1] Jonathan S. Owen,[1] David R. Reichman,[1] and Louis E. Brus[1]*

[1] Department of Chemistry, Columbia University, New York, NY 10027, USA.
[2] Department of Materials and Interfaces, Weizmann Institute of Science, Rehovot, 76100, Israel.
*Corresponding author. Email: leb26@columbia.edu

**Abstract**

We report a strongly temperature dependent luminescence Stokes shift in the electronic spectra of both hybrid and inorganic lead-bromide perovskite single crystals. This behavior stands in stark contrast to that exhibited by more conventional crystalline semiconductors. We correlate the electronic spectra with the anti-Stokes and Stokes Raman vibrational spectra. Dielectric solvation theories, originally developed for excited molecules dissolved in polar liquids, reproduce our experimental observations. Our approach, which invokes a classical Debye-like relaxation process, captures the dielectric response originating from an anharmonic LO phonon at about 20 meV (160 cm$^{-1}$) in the lead-bromide framework. We reconcile the liquid-like picture with more standard solid-state theories of the Stokes shift in crystalline semiconductors.

**Introduction**

Solution processed lead-halide perovskites are highly promising materials for photovoltaic and optoelectronic applications.[1] Despite their inexpensive and facile synthesis, the lead-halide perovskites exhibit long carrier lifetimes and diffusion lengths, as well as low electron-hole recombination rates.[2] These properties have been discussed in terms of Rashba splitting[3,4], ferroelectricity[5,6] and polaron formation[7,8]. This class of materials also exhibits strongly anharmonic atomic displacements at room temperature.[9–12] While marked anharmonicity occurs in some oxide perovskites and other dielectrics,[13] it is unusual in crystalline semiconductors that are typically employed in optoelectronic device applications. Indeed, semiconductor lattice dynamics are typically modeled as harmonic with small perturbative anharmonic corrections.[14]

In this work we examine the difference between the resonance energy in absorption and luminescence (namely, the emission Stokes shift) as a function of temperature (T). We do so for two typical halide perovskites single crystals: hybrid CH$_3$NH$_3$PbBr$_3$ (MA) and all-inorganic CsPbBr$_3$ (Cs). The emission Stokes shift stems from the fact that, to a good approximation, absorption instantaneously excites an exciton resonance before any possible excited state lattice relaxation can occur, while emission probes the inter-band electronic transition after lattice relaxation. We observe that in contrast to conventional IV, III-V and II-VI semiconductors, the emission Stokes shift in lead-halide perovskites

*drastically* increases with T. Using classical dielectric solvation theories, originally developed for describing the Stokes shift of molecules in polar *liquids*, we reproduce the experimental observation. The efficacy of dielectric solvation theory highlights the role of lattice anharmonicity in carrier-lattice interactions in lead-halide perovskites. Yet this result is not in contradiction with the ability of an *effective* harmonic model to describe the emission Stokes shift.

**Results and Discussion**

Figure 1 shows the optical absorption and photoluminescence of single lead bromide perovskite crystals. The absorption coefficient α is calculated from the observed reflectance spectra using a Kramers-Kronig constrained variational analysis. (SI section A and B). Our absorption and emission data agree with that of Tilchin et al[15] who made a careful study of the MA compound below about 200 K. As previously reported, extrinsic photoluminescence side bands due to shallow traps are observed at low temperatures in the MA compound.[15,16] The side bands decrease in magnitude at higher T as thermal de-trapping occurs. Furthermore, near 150 K in MA, an abrupt negative jump in the resonance energy occurs as the orthorhombic to tetragonal phase transition is crossed (gray arrow in figure 1B). Overall, the spectral features of MA and Cs and their temperature evolution are remarkably similar.

Figures 2A and B show the evolution of the peak energy in optical absorption (blue) and photoluminescence (red) for Cs and MA, respectively. These results further emphasize the similar behavior of the hybrid and all-inorganic crystals. The resonance energy of the absorption in both materials shifts to the blue as T increases. This blue shift behavior, related to a positive $dE_g/dT$, is opposite to that of common semiconductors such as GaAs,[17] yet characteristic of lead-based semiconductors.[14]

The resonance energy of the photoluminescence (Figures 2A and B in blue) follows the absorption resonance up to 50-60 K. At higher temperatures the photoluminescence energy diverges from that of the absorption and becomes almost independent of T, while the absorption resonance continues to shift blue as it did at lower T. Heuristically, it appears as if the emission no longer follows the band gap physics seen in the instantaneous absorption. We note that the temperature at which the emission behavior breaks from the absorption does not coincide with any known structural phase transition and is well below the tetragonal-orthorhombic phase transition where organic counter-ions such as MA are known to "freeze-out".[18]

Figures 2C and D show the evolution of the resonance linewidths in optical absorption (blue) and photoluminescence (red) of Cs and MA, respectively. Both absorption and emission resonance linewidths broaden at higher T. Wright et al.[19] showed that emission broadening in the MA compound can be fit by equation 1 at higher T. The first term in equation 1, $\gamma_0$, captures temperature independent broadening factors such as static inhomogeneities and the intrinsic zero-temperature width, while the second term describes broadening from interaction with a thermal population of modes of energy $E_{ph}$ where A is a constant prefactor.[19] The solid curves in figures 2C and 2D show that a phonon energy of about 20 meV reproduces the observed linewidth evolution for both compounds. This observation is in reasonable agreement with that of Wright et al. who observed an activation phonon energy of 15.3 meV (123 cm$^{-1}$). We assign this energy scale to framework LO phonons as did Wright et al.

$$\gamma = \gamma_0 + \frac{A}{\exp\left(\frac{E_{ph}}{kT}\right) - 1} \qquad (1)$$

Both perovskite compounds exhibit similar distributions of framework vibrational modes.[9–11,20] The highest energy optical phonon branch of the lead-bromide lattice occurs at about 20 meV (160 cm$^{-1}$).[21,22] These modes involve predominantly halide octahedral motion.[10,20] Thus we observe the somewhat expected fact that linewidth broadening shows activated behavior due to carrier coupling to LO phonons.

Figure 3A shows the evolution of the emission Stokes shift with temperature for Cs (green dots) and MA (blue dots). Below about 50K the emission Stokes shift is weakly dependent on temperature, and decreases as T increases. This type of Stokes-shifted luminescence is typically produced by potential fluctuations such as compositional disorder, defect sites, and surface states, which allow relaxation and localization into trap sites before radiative recombination.[23] The influence of such extrinsic fluctuations in the emission Stokes shift decreases at higher temperature as detrapping occurs.

From 60 K to 300 K, the emission Stokes shift strongly increases with T. This finding is both surprising and significant. Generally speaking, an intrinsic emission Stokes shift results from lattice relaxation around an optically excited state. For example, in a Frank-Condon picture, the optically excited state prepared by vertical electronic transition subsequently relaxes to lower vibrational states of the excited potential curve, dissipating the excess energy (the Frank-Condon energy, $E_{FC}$) into phonons. For the simplest model of ground and excited state potential energy curves of displaced harmonic oscillators, the Frank-Condon energy is $E_{FC} = S\hbar\omega$, where S is the Huang-Rhys factor and $\hbar\omega$ is the phonon energy.[24] S itself is a geometrical factor with no temperature dependence. In this simple harmonic model the absorption and emission linewidths will broaden as T increases, but the emission Stokes shift is independent of T. This model, often applied to localized states, clearly does not describe our data.

We now seek to understand the anomalous temperature dependence of the emission Stokes shift. Some temperature dependence should come from the T dependence of the static dielectric constant $\varepsilon_0$. Surprisingly, our observed emission Stokes shift is similar to that measured during solvent relaxation around luminescing molecules in polar liquids.[25,26] When a solute molecule is optically excited in a polar solvent, the solvent molecules will rearrange in response to the altered electrostatic configuration of the solute. At low temperatures the solvent is nearly frozen and a small Stokes shift occurs while at higher temperatures, reorientation is facile. The static dielectric constant of a polar liquid thus shows a large increase from low to high temperature. In typical organic dipolar solvents, a dielectric continuum model captures the non-equilibrium solvation around the excited molecule, as described in the Ooshika-Lippert-Mataga (OLM) relation.[27,28] More general descriptions of this dielectric solvation process exist[29] which provide closed-form analytical expressions for the emission Stokes shift $\Delta \tilde{E}(s)$ in the Laplace domain. Here we utilize such models to describe the long time emission Stokes shift via $|\Delta E(\infty) - \Delta E(0)|$.

The dielectric relaxation around a single point charge can be expressed as in equation 2, where $s = i\omega$, a is the radius of a bounding cavity around the solute molecule, and $\varepsilon(s)$ is the dielectric response of the solvent. In our context this is a classical model for relaxation around a point charge carrier. Such a model is reasonable because photoexcitation in perovskites creates nearly free carriers. In this context, the constant a should be no smaller than the excitonic radius.

$$\Delta \tilde{E}(s) = \frac{1}{s}\left(\frac{48}{\pi}\right)^{\frac{1}{3}} \frac{1}{a}\left(1 - \frac{1}{\varepsilon(s)}\right) \qquad (2)$$

$$\varepsilon(s) = \varepsilon_\infty + \frac{\varepsilon_0 - \varepsilon_\infty}{1 + s\tau_D} \qquad (3)$$

A simple description of temperature dependent dielectric response is the Debye dipole relaxation model in equation 3. The importance of the difference of the high and low frequency dielectric constants in determining emission Stokes shift is apparent in both the classical Debye model and the solid state Fan model to be discussed below.

In equation 4 we model the temperature dependence factor in the Debye relaxation, as described in detail in SI section F. Presented in figure 3B is the low frequency limit dielectric function $\varepsilon_0$ obtained by fitting the point charge solvation model to the T dependent emission Stokes shift.

$$\varepsilon_0 - \varepsilon_\infty = \frac{A}{\exp\left(\frac{E_a}{kT}\right) - 1} \qquad (4)$$

For both compounds we find that the parameters A=6.5, a=10nm, and $E_a$=20meV reproduce the data in figure 3A. The extracted value for $E_a$ agrees well with the $E_{ph}$ that was extracted from the linewidth analysis presented in figure 2 following equation 1.

In the dynamics of liquid state processes, it is well-known that the non-equilibrium response of even highly anharmonic systems may be modeled by an effective harmonic model as long as the pertinent fluctuations are Gaussian or if the environment responds in a linear fashion.[30–32] A classical exemplar of this fact is the success of the Marcus theory of electron transfer in polar solvents such as water.[33,34] Indeed, recent simulations demonstrate that even though the perovskite lattice is highly anharmonic, the relevant nuclear fluctuations are statistically nearly Gaussian. Given these facts, we postulate that an effective phononic model of the Stokes shift may be constructed, using the strong temperature dependent dielectric function as input.[35]

From polaron theory, Fan derived equation 5 below for the emission Stokes shift (from the band edge) of recombining free carriers coupled to LO phonons.[36,37] Here $E_{LO}$ is the LO phonon energy, and the other symbols have their usual meaning. This model does not consider excitons. Such a framework works well for GaAs[38], CdSe[39], CdTe[40,41] and other semiconductors at room temperature. In these compounds the emission Stokes shift is just a few meV. Furthermore in such standard cases the static dielectric constant is typically only about 10% larger than the optical dielectric constant.[42] In contrast, the predicted emission Stokes shift (SI section F) for our compounds is large and close to the 60 meV shift we observe; this shift is about three times $E_{LO}$.

$$\Delta E = \left(\frac{1}{\varepsilon_\infty} - \frac{1}{\varepsilon_0}\right)\left(\sqrt{\frac{m_e}{m_0}} + \sqrt{\frac{m_h}{m_0}}\right)\sqrt{|E_{Ryd}|E_{LO}} \qquad (5)$$

It is useful at this stage to compare the classical dielectric solvation model with the polaron of equation 5. In figure 3A, the black dashed line shows the naive prediction of equation 5 with no temperature dependence. The magenta curve shows the prediction of a modified equation 5, valid at all temperatures, using the low frequency dielectric response in equation 4 (Figure 3B). No adjustment is made to the electron/hole mass or the Rydberg energy, nor is any further fitting of dielectric function performed. Both the T dependence and the absolute magnitude of the emission Stokes shift are captured within this model. From this we learn that the strong T dependence of the emission Stokes shift is rooted in a thermally activated large change of the low frequency dielectric constant.

This dielectric response is far stronger than in the more conventional semiconductors. Thus the coupling between the moving carrier and lattice is large in these lead-halide perovskites. Our perovskites show a T dependent dielectric response that behaves roughly as $\varepsilon_0(T) \sim A \times n(T)$ where $A \sim 6$ and n is the Bose-Einstein population. In conventional semiconductors the dielectric response is $\varepsilon_0(T) \sim A \times T$ where $A \sim 10^{-4}$.[42] This difference is not due to different LO phonon frequencies or populations. For example the $E_{LO}$ of CdTe (21 meV) is very similar to that of Cs and MA (~20 meV). A low $E_{LO}$ and the resulting larger phonon population at higher T does not explain the stronger coupling in lead-halide perovskites.

Polarons form when lattice relaxation around carriers becomes significant.[43,44] Large polarons help to explain the stabilization and protection of carriers in lead halide perovskites.[7,8] Scattering by LO phonons is thought to limit the mobility of polarons.[11,45,46] The dielectric response factor $(1/\varepsilon_\infty - 1/\varepsilon_0)$ in equation 5 also appears in the

Frohlich polaron coupling constant. Our findings suggest that in lead-halide perovskites, the carrier-lattice coupling strength should not be considered as a T independent constant.

We hypothesize that this large dielectric response emerges from large amplitude, anharmonic nuclear displacements with weak restoring forces. To generate significant dielectric polarization such displacements are needed. On an anharmonic potential, the dwell time at large displacement can be dominant. In a heuristic sense this can be visualized by comparing the quadratic, Lennard-Jones, and double-well potential curves shown in the inset of Figure 3B. These curves have the same harmonic curvature near their potential minima, yet very different displacements and restoring forces at higher energies.

To examine this hypothesis, we measured the Raman scattering spectra. Figure 4A and D present the temperature dependent (5-90 K) Raman scattering spectra of Cs and MA respectively. Raman modes broaden and shift as they are thermally populated. This thermal population is directly observed in anti-Stokes Raman scattering. Figures 4B and D show that the mode just below 140 $cm^{-1}$ becomes thermally populated near 50-60K, matching the onset of luminescence Stokes shift rise and linewidth broadening discussed above. The framework LO mode energies fall just below 150 $cm^{-1}$ at the Brillouin zone center Γ point.

The Stokes Raman scattering provides further information about phonon lifetime broadening and anharmonicity. Figures 4C and F show that the vibrational modes near 140 $cm^{-1}$ are already significantly broadened near 50 K. In contrast, at the same temperature, the modes at lower frequencies remain sharper, as shown in figures 4A and D. In a Lorentz oscillator model, the Raman mode becomes broad and asymmetric when the damping term is approximately the same magnitude as the resonance energy. Such modes are essentially overdamped oscillators; their quantized motion is not firmly established. The broadening of these modes, at the onset temperature of the emission Stokes shift rise, indicates significant vibrational anharmonicity. Indeed, local polar lattice fluctuations emerging from large amplitude anharmonic nuclear motions have been reported recently.[9,12] Even at liquid nitrogen temperatures, nuclear motion samples the non-harmonic regions of the potential surface leading to lifetime broadening.[9] Broadening and strong anharmonicity have also been observed in the far IR spectra of MA at room temperature.[11] Large amplitude lead-halide octahedral lattice modes responding to photo-generated charge carriers have also been reported on the picosecond to sub-picosecond time scale.[8,47,48]

Summarizing our experimental observations, at low T lattice motions are mostly harmonic (figure 4) and the emission Stokes shift is small (figure 3). At higher T the anharmonic portion of the potential surface is explored and the emission Stokes shift is large. The Bose-Einstein distribution describes vibrational populations in both harmonic and anharmonic curves. In more conventional semiconductors, anharmonicity is treated as a perturbative correction to the harmonic approximation, and Raman spectra remain sharp at higher T.[49] In our compounds anharmonicity is well beyond the perturbative regime. Concomitantly the emergent dielectric response is strong in the presence of charged carriers generated by photoexcitation.

In view of the similar luminescence, optical absorption and Raman data of the Cs and MA compounds, strong dielectric solvation is inherent to an anharmonic lead-halide perovskite framework. This result is supported by theory as discussed above. Yet, note that specific nature of the A site cation does have a clear influence on the abruptness of phase transitions, the microwave and radio frequency range dielectric response, and hot carrier relaxation.[50–53]

**Conclusion**

In conclusion, we report that luminescing band edge carriers in Cs and MA above liquid nitrogen temperatures are stabilized by dielectric response from thermally-activated anharmonic framework modes. The carrier-lattice coupling strength increases at higher temperatures. This creates a strongly temperature dependent emission Stokes shift. This behavior is anomalous compared with more conventional crystalline semiconductors. A thermally-activated Debye liquid dielectric solvation model provides insight into this process, and can be used to predict the temperature dependence, as can an effective polaron model for coupling to LO phonons. Vibrational Raman spectra and prior theoretical modeling indicate the dielectric solvation is provided by the halide group modes near 20 meV. A similar dielectric response is observed in both Cs and MA. This suggests that a large dielectric response from strongly anharmonic lattice dynamics is intrinsic to the lead halide perovskite family.

**Figures and Tables**

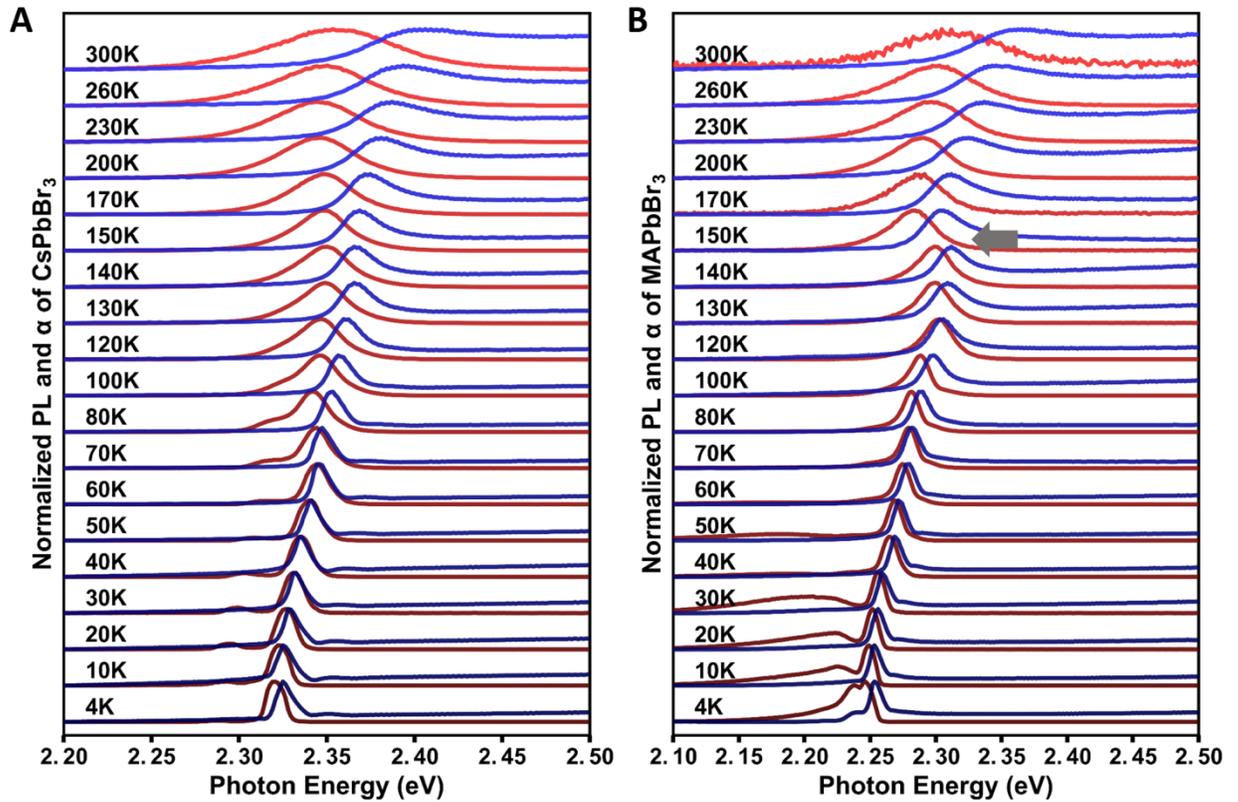

**Fig. 1. Absorption and photoluminescence from 4K to 300K.** A) Cs, B) MA. Blue lines show absorption coefficient. Red lines show photoluminescence. Absorption coefficient and photoluminescence spectra at each temperature are scaled and offset vertically for display clarity. Gray arrow indicates the orthorhombic-tetragonal phase transition in MA.

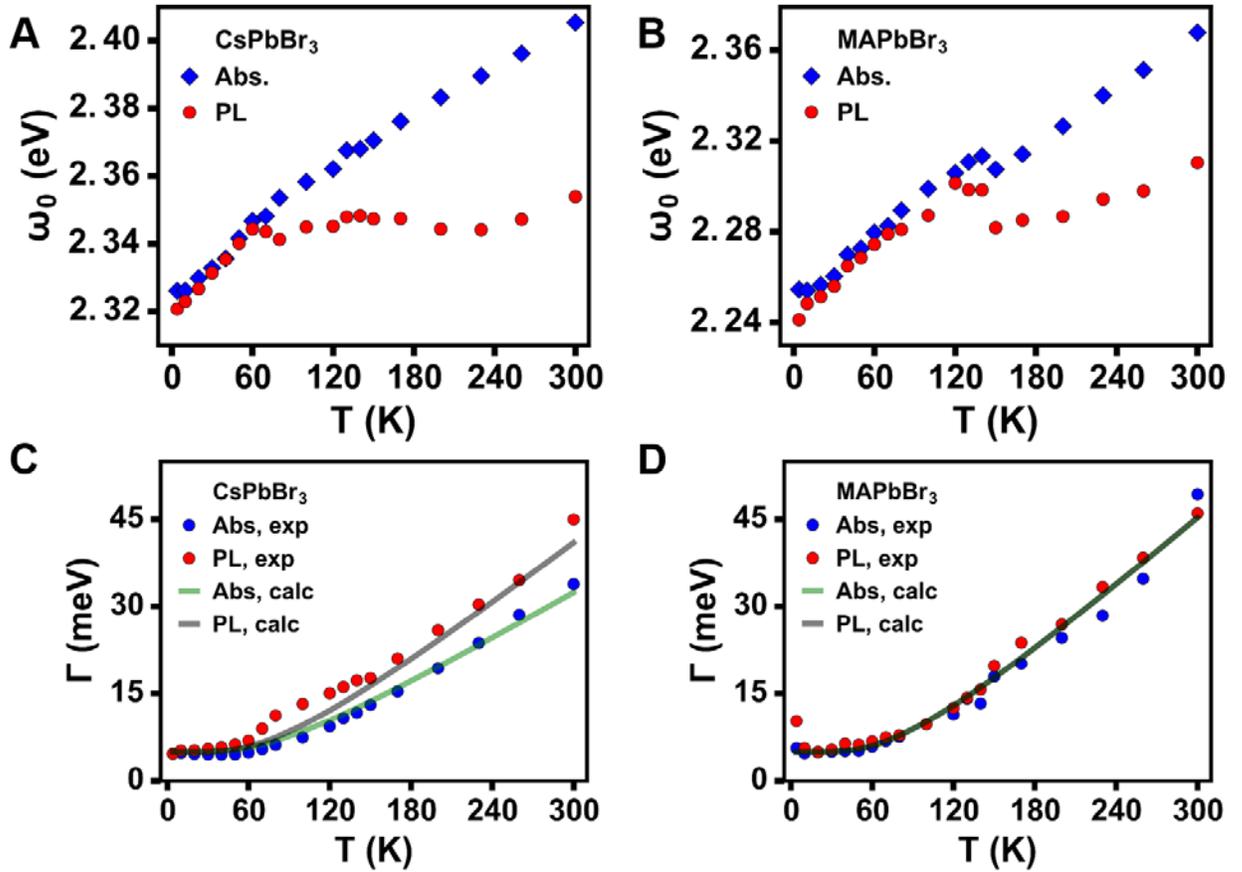

**Fig. 2. Evolution of spectral features of Cs and MA with temperature.** A-B) Absorption and emission resonance energies of Cs (A) and MA (B). C-D) Absorption and emission linewidths of Cs (C) and MA (D). For both materials, blue dots denote the absorption coefficient, α, and red dots denote luminescence energies. Green and gray curves show the linewidth evolution predicted by equation 1, in which $E_{ph}$ = 20meV, and $\gamma_0$ = 5meV. For Cs, A = 32 for absorption and A = 42 for luminescence. For MA, A = 47 for both absorption and luminescence.

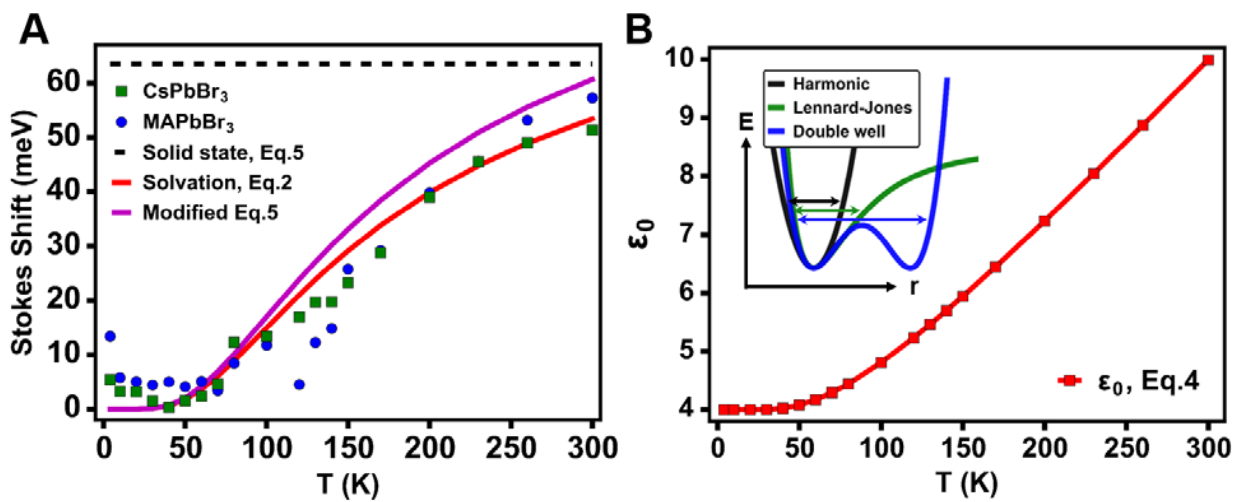

**Fig. 3. Temperature dependent emission Stokes shift and dielectric response.**
A) Emission Stokes shift of Cs (green squares) and MA (blue dots). The two emission Stokes shifts are similar despite their differences in resonance energy and phase sequence. Black dashed curve shows the naive prediction from the Fan equation 5 with constant $\varepsilon_0$. The red curve shows the prediction the dielectric solvation model equation 2. Magenta curve shows the revised prediction from the Fan model using $\varepsilon_0(T)$ from figure 3B. The insert in 3B shows harmonic, Lennard-Jones, and double well schematic potentials. All three curves share the same energy and curvature near the potential minima.

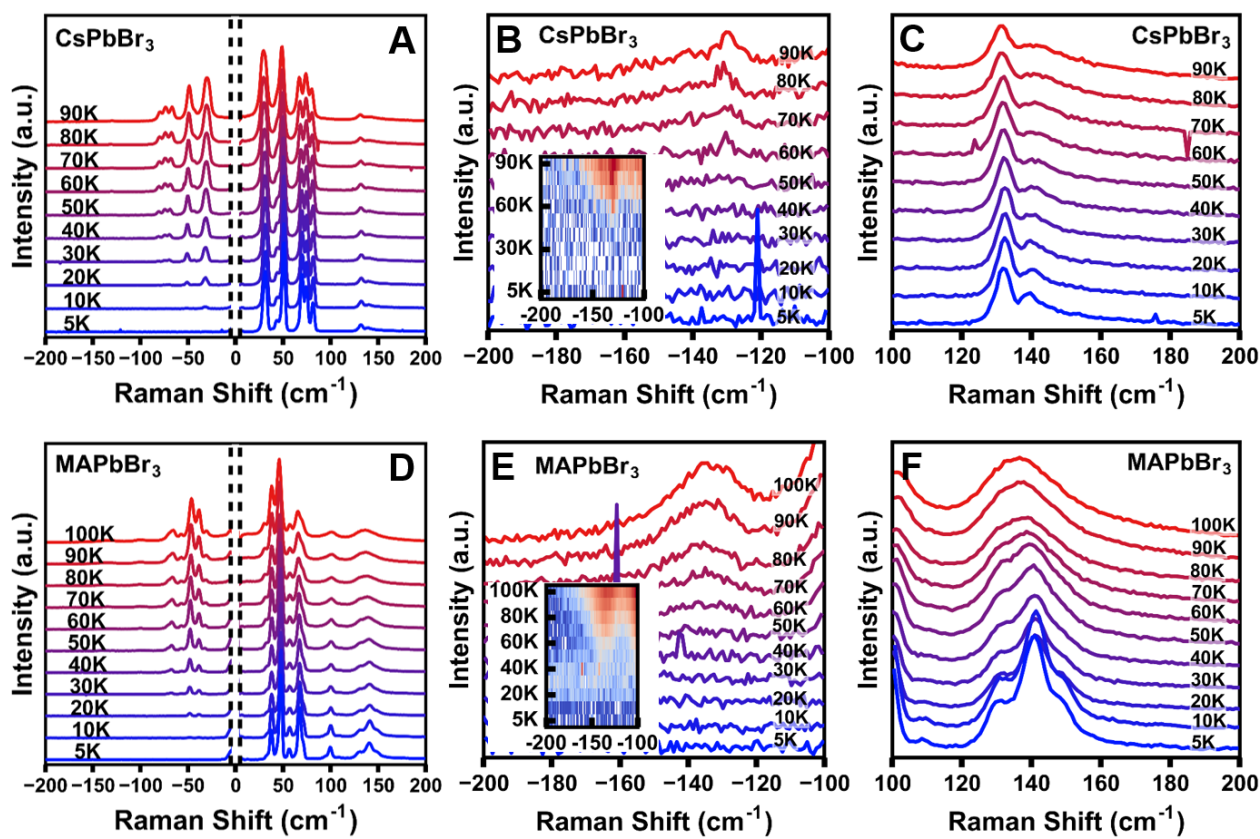

**Fig. 4. Temperature dependent Raman scattering of Cs and MA.** A) Cs. B) Anti-Stokes and C) Stokes Raman scattering of the Cs framework mode just below 150 cm-1. D) MA. E) Anti-Stokes and F) Stokes Raman scattering of the MA framework modes just below 150 cm-1. Insets in B) and E) show the corresponding spectral datasets as color plots.

## References

(1) Stranks, S. D.; Snaith, H. J. Metal-Halide Perovskites for Photovoltaic and Light-Emitting Devices. *Nature Nanotechnology* **2015**, *10*, 391.

(2) Brenner, T. M.; Egger, D. A.; Kronik, L.; Hodes, G.; Cahen, D. Hybrid Organic—inorganic Perovskites: Low-Cost Semiconductors with Intriguing Charge-Transport Properties. *Nature Reviews Materials* **2016**, *1*, 15007.

(3) Zheng, F.; Tan, L. Z.; Liu, S.; Rappe, A. M. Rashba Spin–Orbit Coupling Enhanced Carrier Lifetime in CH3NH3PbI3. *Nano Lett.* **2015**, ASAP.

(4) Isarov, M.; Tan, L. Z.; Bodnarchuk, M. I.; Kovalenko, M. V; Rappe, A. M.; Lifshitz, E. Rashba Effect in a Single Colloidal CsPbBr3 Perovskite Nanocrystal Detected by Magneto-Optical Measurements. *Nano Letters* **2017**, *17* (8), 5020–5026.

(5) Rakita, Y.; Bar-Elli, O.; Meirzadeh, E.; Kaslasi, H.; Peleg, Y.; Hodes, G.; Lubomirsky, I.; Oron, D.; Ehre, D.; Cahen, D. Tetragonal CH3NH3PbI3 Is Ferroelectric. *Proceedings of the National Academy of Sciences* **2017**, *114* (28), E5504–E5512.

(6) Rakita, Y.; Meirzadeh, E.; Bendikov, T.; Kalchenko, V.; Lubomirsky, I.; Hodes, G.; Ehre, D.; Cahen, D. CH3NH3PbBr3 Is Not Pyroelectric, Excluding Ferroelectric-Enhanced Photovoltaic Performance. *APL Materials* **2016**, *4* (5), 51101.

(7) Zhu, X.-Y.; Podzorov, V. Charge Carriers in Hybrid Organic–Inorganic Lead Halide Perovskites Might Be Protected as Large Polarons. *J. Phys. Chem. Lett.* **2015**, *6* (23), 4758–4761.

(8) Miyata, K.; Meggiolaro, D.; Trinh, M. T.; Joshi, P. P.; Mosconi, E.; Jones, S. C.; De Angelis, F.; Zhu, X.-Y. Large Polarons in Lead Halide Perovskites. *Science Advances* **2017**, *3* (8).

(9) Yaffe, O.; Guo, Y.; Tan, L. Z.; Egger, D. A.; Hull, T.; Stoumpos, C. C.; Zheng, F.; Heinz, T. F.; Kronik, L.; Kanatzidis, M. G.; et al. Local Polar Fluctuations in Lead Halide Perovskite Crystals. *Phys. Rev. Lett.* **2017**, *118* (13), 136001.

(10) Leguy, A. M. A.; Goni, A. R.; Frost, J. M.; Skelton, J.; Brivio, F.; Rodriguez-Martinez, X.; Weber, O. J.; Pallipurath, A.; Alonso, M. I.; Campoy-Quiles, M.; et al. Dynamic Disorder, Phonon Lifetimes, and the Assignment of Modes to the Vibrational Spectra of Methylammonium Lead Halide Perovskites. *Phys. Chem. Chem. Phys.* **2016**, *18* (39), 27051–27066.

(11) Sendner, M.; Nayak, P. K.; Egger, D. A.; Beck, S.; Muller, C.; Epding, B.; Kowalsky, W.; Kronik, L.; Snaith, H. J.; Pucci, A.; et al. Optical Phonons in Methylammonium Lead Halide Perovskites and Implications for Charge Transport. *Mater. Horiz.* **2016**, *3* (6), 613–620.

(12) Guo, P.; Xia, Y.; Gong, J.; Stoumpos, C. C.; McCall, K. M.; Alexander, G. C. B.; Ma, Z.; Zhou, H.; Gosztola, D. J.; Ketterson, J. B.; et al. Polar Fluctuations in Metal Halide Perovskites Uncovered by Acoustic Phonon Anomalies. *ACS Energy Letters* **2017**, 2463–2469.

(13) Scott, J. F. Soft-Mode Spectroscopy: Experimental Studies of Structural Phase Transitions. *Rev. Mod. Phys.* **1974**, *46* (1), 83–128.

(14) Ashcroft, N. W.; Mermin, N. D. *Solid State Physics*; Holt, Rinehart and Winston: New

York.

(15) Tilchin, J.; Dirin, D. N.; Maikov, G. I.; Sashchiuk, A.; Kovalenko, M. V.; Lifshitz, E. Hydrogen-like Wannier-Mott Excitons in Single Crystal of Methylammonium Lead Bromide Perovskite. *ACS Nano* **2016**, *10* (6), 6363–6371.

(16) Lozhkina, O. A.; Yudin, V. I.; Murashkina, A. A.; Shilovskikh, V. V.; Davydov, V. G.; Kevorkyants, R.; Emeline, A. V.; Kapitonov, Y. V.; Bahnemann, D. W. Low Inhomogeneous Broadening of Excitonic Resonance in MAPbBr 3 Single Crystals. *The Journal of Physical Chemistry Letters* **2018**, 302–305.

(17) Fox, M. *Optical Properties of Solids*; Oxford university press, 2010; Vol. 3.

(18) Baikie, T.; Barrow, N. S.; Fang, Y.; Keenan, P. J.; Slater, P. R.; Piltz, R. O.; Gutmann, M.; Mhaisalkar, S. G.; White, T. J. A Combined Single Crystal neutron/X-Ray Diffraction and Solid-State Nuclear Magnetic Resonance Study of the Hybrid Perovskites CH3NH3PbX3 (X = I, Br and Cl). *Journal of Materials Chemistry A* **2015**, *3* (17), 9298–9307.

(19) Wright, A. D.; Verdi, C.; Milot, R. L.; Eperon, G. E.; Pérez-Osorio, M. A.; Snaith, H. J.; Giustino, F.; Johnston, M. B.; Herz, L. M. Electron–phonon Coupling in Hybrid Lead Halide Perovskites. *Nature Communications* **2016**, *7* (May).

(20) Guo, Y.; Yaffe, O.; Paley, D. W.; Beecher, A. N.; Hull, T. D.; Szpak, G.; Owen, J. S.; Brus, L. E.; Pimenta, M. A. Interplay between Organic Cations and Inorganic Framework and Incommensurability in Hybrid Lead-Halide Perovskite CH3NH3PbBr3. *Phys. Rev. Materials* **2017**, *1* (4), 42401.

(21) Huang, L.; Lambrecht, W. R. L. Lattice Dynamics in Perovskite Halides CsSnX3 with X = I, Br, Cl. *Physical Review B* **2014**, *90* (19), 195201.

(22) da Silva, E. L.; Skelton, J. M.; Parker, S. C.; Walsh, A. Phase Stability and Transformations in the Halide Perovskite CsSnI3. *Phys. Rev. B* **2015**, *91* (14), 144107.

(23) Yang, F.; Wilkinson, M.; Austin, E. J.; O'Donnell, K. P. Origin of the Stokes Shift: A Geometrical Model of Exciton Spectra in 2D Semiconductors. *Phys. Rev. Lett.* **1993**, *70* (3), 323–326.

(24) de Jong, M.; Seijo, L.; Meijerink, A.; Rabouw, F. T. Resolving the Ambiguity in the Relation between Stokes Shift and Huang-Rhys Parameter. *Phys. Chem. Chem. Phys.* **2015**, *17* (26), 16959–16969.

(25) Arzhantsev, S.; Jin, H.; Baker, G. A.; Maroncelli, M. Measurements of the Complete Solvation Response in Ionic Liquids. *The Journal of Physical Chemistry B* **2007**, *111* (18), 4978–4989.

(26) Kashyap, H. K.; Biswas, R. Stokes Shift Dynamics in Ionic Liquids: Temperature Dependence. *Journal of Physical Chemistry B* **2010**, *114* (50), 16811–16823.

(27) Mataga, N.; Kaifu, Y.; Koizumi, M. The Solvent Effect on Fluorescence Spectrum, Change of Solute-Solvent Interaction during the Lifetime of Excited Solute Molecule. *Bulletin of the Chemical Society of Japan* **1955**, *28* (9), 690–691.

(28) Bagchi, B.; Oxtoby, D. W.; Fleming, G. R. Theory of the Time Development of the Stokes Shift in Polar Media. *Chemical Physics* **1984**, *86* (3), 257–267.

(29) Song, X.; Chandler, D.; Marcus, R. A. Gaussian Field Model of Dielectric Solvation

Dynamics. *J. Phys. Chem.* **1996**, *100* (29), 11954–11959.

(30) Bader, J. S.; Kuharski, R. A.; Chandler, D. Role of Nuclear Tunneling in Aqueous Ferrous–ferric Electron Transfer. *The Journal of Chemical Physics* **1990**, *93* (1), 230–236.

(31) Bader, J. S.; Chandler, D. Computer Simulation of Photochemically Induced Electron Transfer. *Chemical Physics Letters* **1989**, *157* (6), 501–504.

(32) Hwang, J. K.; Warshel, A. Microscopic Examination of Free-Energy Relationships for Electron Transfer in Polar Solvents. *Journal of the American Chemical Society* **1987**, *109* (3), 715–720.

(33) Marcus, R. A. Electron Transfer Reactions in Chemistry. Theory and Experiment. *Rev. Mod. Phys.* **1993**, *65* (3), 599–610.

(34) Kuharski, R. A.; Bader, J. S.; Chandler, D.; Sprik, M.; Klein, M. L.; Impey, R. W. Molecular Model for Aqueous Ferrous–ferric Electron Transfer. *The Journal of Chemical Physics* **1988**, *89* (5), 3248–3257.

(35) Mayers, M. Z.; Fan, Z.; Egger, D. A.; Rappe, A. M.; Reichman, D. R. To Be Submitted. **2018**.

(36) Ullrich, B.; Singh, A. K.; Barik, P.; Xi, H.; Bhowmick, M. Inherent Photoluminescence Stokes Shift in GaAs. *Optics Letters* **2015**, *40* (11), 2580–2583.

(37) Fan, H. Y. Photon-Electron Interaction, Crystals Without Fields BT - Light and Matter Ia / Licht Und Materie Ia. In; Genzel, L., Ed.; Springer Berlin Heidelberg: Berlin, Heidelberg, 1967; pp 157–233.

(38) Sasaki, C.; Naito, H.; Iwata, M.; Kudo, H.; Yamada, Y.; Taguchi, T.; Jyouichi, T.; Okagawa, H.; Tadatomo, K.; Tanaka, H. Temperature Dependence of Stokes Shift in $In_xGa_{1-x}N$ Epitaxial Layers. *Journal of Applied Physics* **2003**, *93* (3), 1642–1646.

(39) Liptay, T. J.; Marshall, L. F.; Rao, P. S.; Ram, R. J.; Bawendi, M. G. Anomalous Stokes Shift in CdSe Nanocrystals. *Phys. Rev. B* **2007**, *76* (15), 155314.

(40) Morikawa, K.; Yamaka, E. Temperature Dependence of the Electroluminescence in CdTe Diodes. *Japanese Journal of Applied Physics* **1968**, *7* (3), 243.

(41) Watanabe, T.; Takahashi, K.; Shimura, K.; Kim, D. Influence of Carrier Localization at the Core/shell Interface on the Temperature Dependence of the Stokes Shift and the Photoluminescence Decay Time in CdTe/CdS Type-II Quantum Dots. *Phys. Rev. B* **2017**, *96* (3), 35305.

(42) Strzalkowski, I.; Joshi, S.; Crowell, C. R. Dielectric Constant and Its Temperature Dependence for GaAs, CdTe, and ZnSe. *Applied Physics Letters* **1976**, *28* (6), 350–352.

(43) Mott, N. F.; Stoneham, A. M. The Lifetime of Electrons, Holes and Excitons before Self-Trapping. *Journal of Physics C: Solid State Physics* **1977**, *10* (17), 3391.

(44) Stoneham, A. M.; Gavartin, J.; Shluger, A. L.; Kimmel, A. V; Ramo, D. M.; Rønnow, H. M.; Aeppli, G.; Renner, C. Trapping, Self-Trapping and the Polaron Family. *Journal of Physics: Condensed Matter* **2007**, *19* (25), 255208.

(45) Frost, J. M. Calculating Polaron Mobility in Halide Perovskites. *Phys. Rev. B* **2017**, *96* (19), 195202.

(46) Zhang, M.; Zhang, X.; Huang, L.-Y.; Lin, H.-Q.; Lu, G. Charge Transport in Hybrid Halide Perovskites. *Phys. Rev. B* **2017**, *96* (19), 195203.

(47) Wu, X.; Tan, L. Z.; Shen, X.; Hu, T.; Miyata, K.; Trinh, M. T.; Li, R.; Coffee, R.; Liu, S.; Egger, D. A.; et al. Light-Induced Picosecond Rotational Disordering of the Inorganic Sublattice in Hybrid Perovskites. *Science Advances* **2017**, *3* (7).

(48) Batignani, G.; Fumero, G.; Kandada, A. R. S.; Cerullo, G.; Gandini, M.; Ferrante, C.; Petrozza, A.; Scopigno, T. Probing Femtosecond Lattice Displacement upon Photo-Carrier Generation in Lead Halide Perovskite. *arXiv* **2017**.

(49) Cowley, R. A. Anharmonic Crystals. *Reports on Progress in Physics* **1968**, *31* (1), 123.

(50) Poglitsch, A.; Weber, D. Dynamic Disorder in Methylammoniumtrihalogenoplumbates (II) Observed by Millimeter-wave Spectroscopy. *The Journal of Chemical Physics* **1987**, *87* (11), 6373.

(51) Onoda-Yamamuro, N.; Matsuo, T.; Suga, H. Dielectric Study of CH3NH3PbX3 (X = Cl, Br, I). *Journal of Physics and Chemistry of Solids* **1992**, *53* (7), 935–939.

(52) Zhu, H.; Miyata, K.; Fu, Y.; Wang, J.; Joshi, P. P.; Niesner, D.; Williams, K. W.; Jin, S.; Zhu, X.-Y. Screening in Crystalline Liquids Protects Energetic Carriers in Hybrid Perovskites. *Science* **2016**, *353* (6306), 1409–1413.

(53) Anusca, I.; Balčiūnas, S.; Gemeiner, P.; Svirskas, Š.; Sanlialp, M.; Lackner, G.; Fettkenhauer, C.; Belovickis, J.; Samulionis, V.; Ivanov, M.; et al. Dielectric Response: Answer to Many Questions in the Methylammonium Lead Halide Solar Cell Absorbers. *Advanced Energy Materials* **2017**.

**Supporting Information:**

**Liquid-like Free Carrier Solvation and Band Edge Luminescence in Lead-Halide Perovskites**

Yinsheng Guo,[1] Omer Yaffe,[2] Trevor D. Hull,[1] Jonathan S. Owen,[1] David R. Reichman,[1] and Louis E. Brus[1]*

[1] Department of Chemistry, Columbia University, New York, NY 10027, USA.
[2] Department of Materials and Interfaces, Weizmann Institute of Science, Rehovot, 76100, Israel.
*Corresponding author. Email: leb26@columbia.edu

**Materials and Methods**

*Single crystal synthesis.* High quality, mm-sized single crystals of MA and Cs were synthesized as reported previously.[1,2] Briefly, $CH_3NH_2$ (40% w/w Alfa Aesar) and HBr (48% Acros) were mixed to prepare $CH_3NH_3Br$. $PbBr_2$ (98% Sigma-Aldrich) and $CH_3NH_3Br$ were then dissolved in N,N-dimethylformamide at 1:1 ratio to form 1M solution. Single crystals of MA were grown by diffusion of isopropyl alcohol vapor into the solution. Single crystals of Cs were grown by diffusion of n-propyl alcohol vapor into a 30 mM solution of 1:1 lead bromide (98%, Sigma-Aldrich) and cesium bromide (99% metal basis, Alfa Aesar) in N,N- dimethylformamide.

*Low temperature optical measurements.* Single crystals were mounted onto the cold finger of an optical cryostat. The sample chamber was pumped down to the order of $10^{-5}$ torr, before cooling down with liquid Helium.

*Optical reflectance spectroscopy.* A quartz-tungsten-halogen lamp was used as the white light source. The white light passed through a 100 micron pinhole and was collimated. The white light beam was then sent into an inverted microscope, and imaged onto the back aperture of a 40x long working distance objective with NA=0.6. The objective focused the beam onto samples at normal incidence. The reflected light was collected by the same objective and imaged onto the entrance slit of a spectrograph. A 150 grooves/mm grating dispersed the reflected light on a thermoelectrically cooled CCD detector. For each measurement, reflectance of fused silica was collected under the same conditions, which was used as a reference.[3]

*Photoluminescence spectroscopy.* Photoluminescence was excited with the 458nm line of an Ar ion laser. The excitation laser was focused on samples using a 40x objective with NA=0.6. Photoluminescence was collected by the same objective, imaged onto the spectrograph entrance slit, and dispersed by a 150 grooves/mm grating on the CCD detector. Typical excitation power was about 0.2 W/cm$^2$ at low temperatures (T<200 K), and about 1 W/cm$^2$ at higher temperatures (T>200 K).

*Low-frequency Raman spectroscopy and temperature calibration.* The 633 nm line of a Helium-Neon laser was used as the Raman excitation. Spurious light in the laser beam was rejected by a 90/10 volume holographic grating (VHG) beam splitter. The beam was coupled into a home-built inverted microscope, and focused on samples with a 40x objective with NA=0.6. The Raman scattered light was collected by the same objective. Rayleigh scattered light was rejected first by the 90/10 VHG beam splitter, and then by two VHG notch filters. Each notch filter has OD>4 rejection, and a spectral cutoff full width half maximum about 7 $cm^{-1}$ around the 633 nm laser line. After passing through the VHG filters, the Raman signal was spatially filtered, imaged onto the entrance slit of a spectrograph, dispersed by a 1800 grooves/mm grating onto a liquid nitrogen cooled Si CCD detector. Spectral collection covered anti-Stokes and Stokes region from -200 $cm^{-1}$ to 200 $cm^{-1}$. The anti-Stokes to Stokes ratio was used to calculate sample spot temperature as expected from the Bose-Einstein distribution. The cryostat temperature control and readout were found to be accurate within ±1K. Data analysis was done using the Python programming language. Both reflectance and photoluminescence data was fit with the lmfit package.[4]

### A. Analysis of reflectance and photoluminescence spectra

Figure S1 shows optical reflectance and photoluminescence of Cs and MA as a function of temperature.

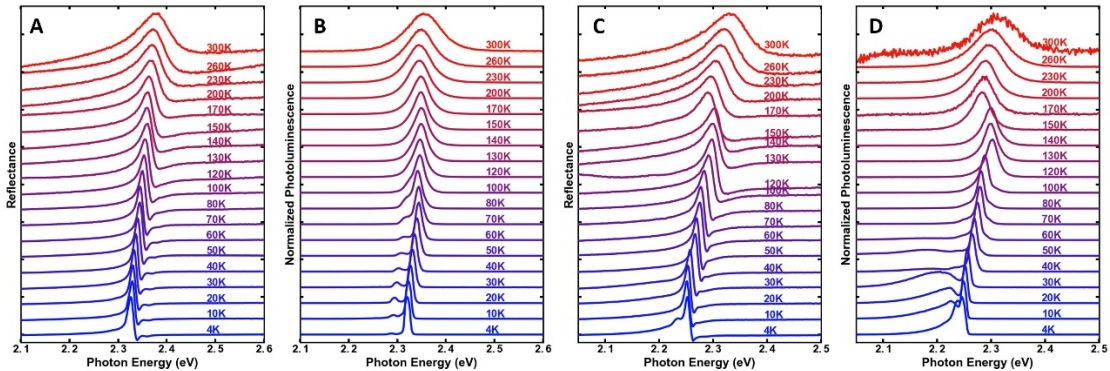

**Fig. S1. Optical reflectance and photoluminescence spectra of Cs and MA from 4K to 300K.** A. Cs optical reflectance, B. Cs photoluminescence, C. MA optical reflectance, D. MA photoluminescence. Each photoluminescence spectrum is normalized with respect to its peak intensity. Reflectance and photoluminescence spectra are vertically offset for display clarity.

### B. Kramers-Kronig analysis of reflection.

We calculated the absorption coefficient from the measured reflectance, using Kramers-Kronig constrained variational analysis.[5–7] The key to the calculation of the absorption coefficient was determining the complex dielectric functions using reflectance data, as expressed in equations S1-S3. To extract the complex dielectric function, we used a sum of Lorentz oscillators and a constant background, as expressed in equation S4. The use of Lorentz oscillator functional form guarantees that the Kramers-Kronig relationship is valid.

$$R(\omega) = \frac{[n(\omega) - 1]^2 + k(\omega)^2}{[n(\omega) + 1]^2 + k(\omega)^2} \quad (S1)$$

$$\alpha(\omega) = \frac{2\omega}{c} k(\omega) \quad (S2)$$

$$n + ik = \sqrt{\varepsilon} \quad (S3)$$

$$\varepsilon = \varepsilon_\infty + \sum_{k=1}^{N} \frac{\omega_{p,k}^2}{\omega_{0,k}^2 - \omega^2 - i\omega\gamma_k} \quad (S4)$$

Numerically, we used N = 256 Lorentz oscillators, half the number of our available data points in each spectrum. The oscillators were placed at fixed positions, evenly spaced out with 3 meV intervals. The oscillators' width were fixed with $\gamma_k$ = 5 meV. The oscillator amplitudes and background level were adjustable parameters. This numerical procedure produced a robust representation of the reflectance data.

In figure S2, we show an example of the process, including reflectance spectrum, initial guess values of a regularly spaced, dense forest of peaks, and final spectrum from fitting. Dielectric functions were obtained from the fitting results, according to equation S4. Complex refractive indices and absorption coefficient were obtained, according to equation S3 and equation S2.

We note that, for accurate determination of the optical dielectric function, one has to take into account the contributions from high- and low-lying transitions beyond the measured spectral range. In the fitting the high energy end of the dielectric function showed an unphysical down turn. The complex refractive indices showed the same behavior accordingly. This is due to the absence of higher energy resonances beyond the spectral range of our data. This apparent lineshape distortion is excluded in subsequent data analysis.

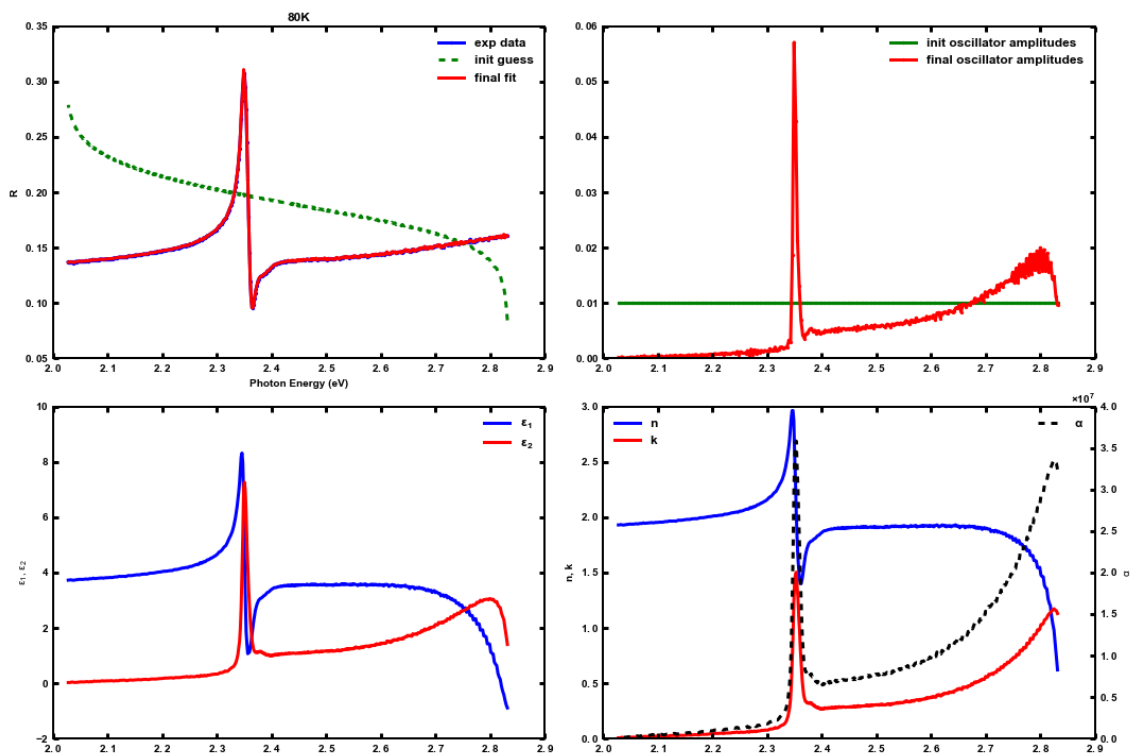

**Fig. S2. Example of Kronig-Kramers constrained variational analysis.** Example spectrum was reflectance of Cs at 80K.

To extract spectral characteristics of the resonance, we fit both the absorption coefficient and photoluminescence resonances with a single Lorentzian function. Spectral background features other than the resonance were modelled with a cubic baseline for absorption and a constant baseline for photoluminescence.

### C. Dielectric function in the optical frequency range

Optical frequency dielectric functions are obtained from the Kramers-Kronig constrained variational analysis as described above. In figure S3, we plot the dielectric function below the main resonance at the low energy end of our spectral range. The dielectric function in this range is largely constant, independent of temperature evolution. Thus we use $\varepsilon_\infty = 4$ for our subsequent analysis of dielectric solvation.

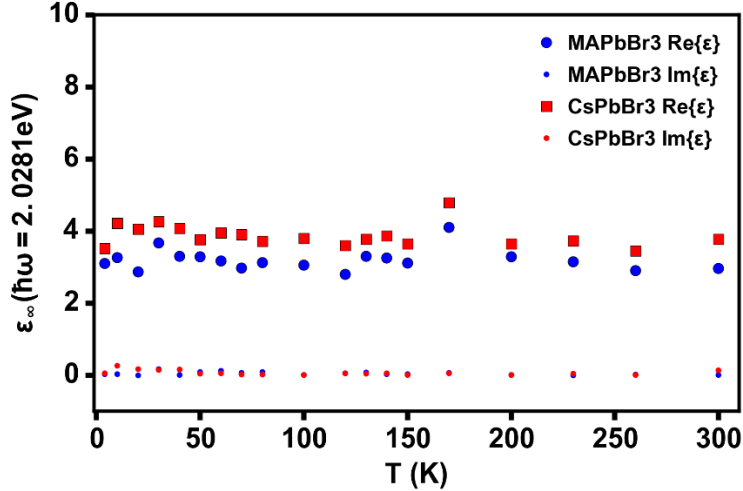

**Fig. S3. Temperature dependence of optical dielectric function below bandgap.**

### D. Prediction of emission Stokes shift based on main text equation 5

In the naive prediction of emission Stokes shift based on the solid state Fan model equation 5 in the main text, we used the following parameters:

$m_0$: mass of free electron.
$E_{Ryd}$: Rydberg energy of hydrogen atom in vacuum, 13.6 eV.
$m_e$: band mass of electron, $0.13 m_0$ taken from Miyata et al.[8]
$m_h$: band mass of hole, $0.19 m_0$ taken from Miyata et al.[8]
$E_{LO}$: the LO phonon energy, 20 meV as obtained from the fitting of emission Stokes shift as well as the fitting of absorption and emission linewidth.
$\varepsilon_\infty$: dielectric constant at the high frequency limit, 4.0 as obtained from our optical reflectance measurements.
$\varepsilon_0$: dielectric constant at the low frequency limit, 10.75 as taken from Tilchin et al.[9]

### E. Numerical inversion of Laplace transform

The numerical inversion of a Laplace transform is widely performed.[10] Here we implement the Talbot and the Gaver-Stehfast algorithms. Using simple analytical test function pairs, we test the validity of the inverse transform. As shown in figure S4 A to D, for exponential, quadratic, linear and constant functions, both algorithms return valid results, identical within the numerical precision. As shown in figure S4 E and F, for oscillatory functions both Talbot and Stehfast algorithms fail, and return results that deviate from true values and differ from each other. The occurrence of differences in the inverse transform output can be used to indicate numerical failure and serve as an inherent consistency check.

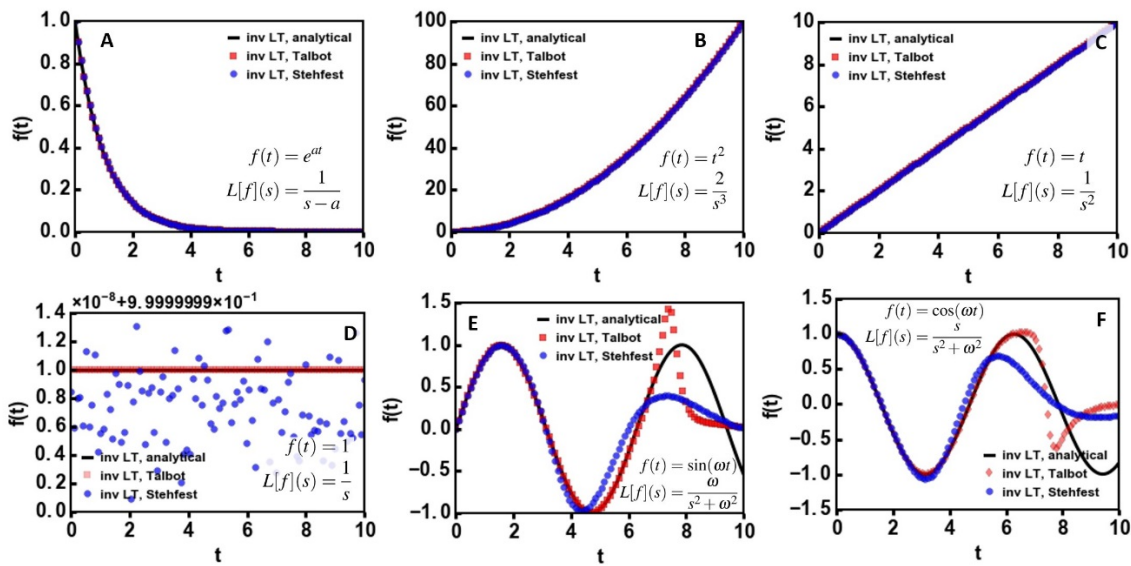

**Fig. S4. Numerical behaviors of inverse Laplace transform algorithms on simple analytical functions.**

Next we use the Talbot and the Gaver-Stehfest algorithm to obtain the inverse Laplace transform of the Gaussian field model. The two algorithms are applied to the dielectric solvation processes of a point charge and a point dipole, expressed in equation S5 and S8. The results are shown in figure S5. Both algorithms return the same result in the two scenarios. Moreover, at the limit of very large and very small s in the Laplace domain, the Debye relaxation model reduces to the limiting values. The numerical inversion of the Gaussian field model returns the corresponding limiting results consistently.

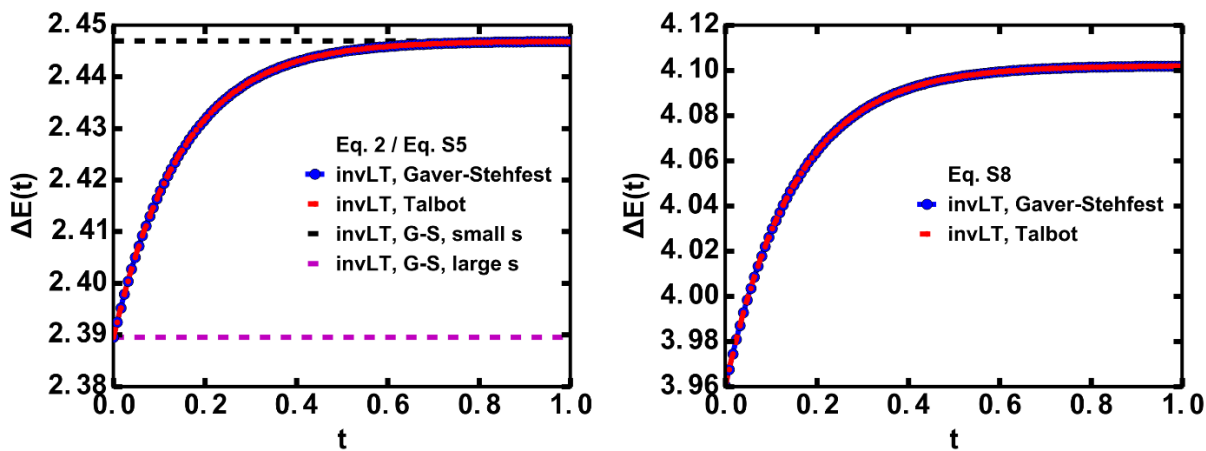

**Fig. S5. Inverse Laplace transform of dielectric solvation dynamics for point charge and point dipole in a dielectric continuum.**

Having obtained the dielectric solvation dynamics in the time domain numerically, we examine how the dielectric response affects solvation dynamics. Using the Debye relaxation model expressed in the main text with the point charge solvation, we alter the parameters $\varepsilon_\infty$, $\varepsilon_0$, and $\tau$ one at a time and study the effect of each parameter. The results are shown in figure S6. The parameter $\tau$ determines the time constant of the relaxation process. These changes do not affect the extent and strength of screening, only the time needed to reach steady state. The parameter $\varepsilon_\infty$ describes the high frequency, or equivalently short time scale, dielectric response. This affects the screening dynamics shortly after excitation, but not the steady state. As $\varepsilon_\infty$ increases, $|\Delta E(t = 0)|$ increases, and $|\Delta E(t = \infty)|$ remains constant. The parameter $\varepsilon_0$ describes the low frequency, or equivalently long time scale dielectric response. This affects the screening dynamics long after excitation. As $\varepsilon_0$ increases, $|\Delta E(t = \infty)|$ increases, and $|\Delta E(t = 0)|$ remains constant. We note that this examination treats $\varepsilon_0$, $\varepsilon_\infty$, and $\tau$ effectively as independent variables. In some physically driven models, $\varepsilon_0$ and/or $\varepsilon_\infty$ are often related to $\tau$.[11] The above examination shows that, $\varepsilon_0 - \varepsilon_\infty$ in the numerator is the main factor affecting the steady-state Stokes shift, $\tau$ in the denominator has negligible influence on the steady-state Stokes shift. When $\varepsilon_0 - \varepsilon_\infty$ is temperature dependent, the observed steady-state Stoke shift will thus change as a function of temperature.

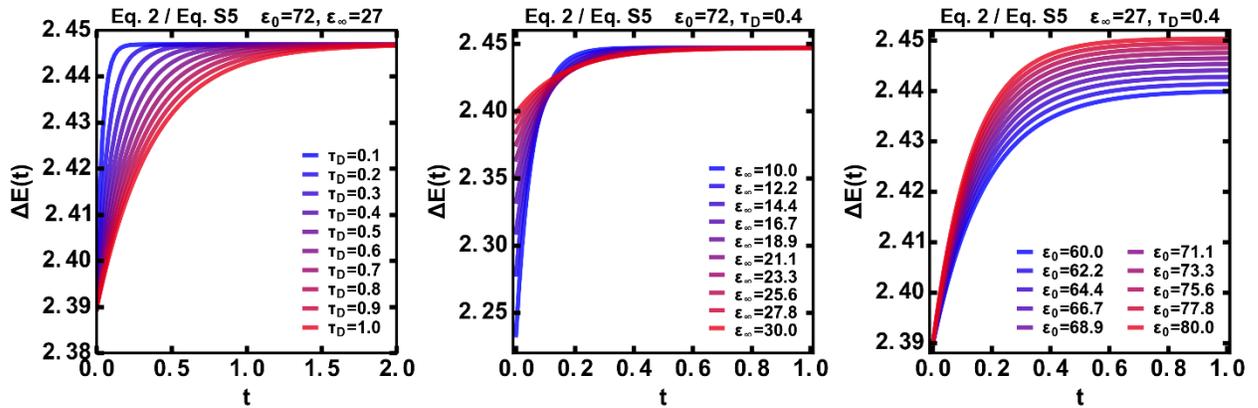

**Fig. S6. Effect of Debye relaxation parameters on the dynamics of dielectric solvation.**

**F. Gaussian field solvation model fitting of the T dependent Stokes shift and Dielectric function $\varepsilon_0$**

*Point charge solvation model for fitting the T dependent Stokes shift*

As discussed in the text, the Gaussian solvation model provides an analytical solution of the solvation dynamics in the Laplace domain, as shown in equation S5 (equation 2 in the main text). Here a is the solvation cell radius, $\varepsilon(s)$ is the dielectric function, and $s = i\omega$. The dielectric response function of the solvent can be described by a Debye relaxation model, as shown in equation S6, where $\tau_D$ is the characteristic time scale of dielectric relaxation. $\tau_D$ is sometimes modeled as a function of T and related to $\varepsilon_0 - \varepsilon_\infty$,[11] but it is not always necessary.[12] $\tau_D$ as an independent variable is shown to be insignificant in determining steady-state Stokes shift, as discussed in section E. The temperature dependence of dielectric response can be modelled with equation S7.

$$\Delta \tilde{E}(s) = \frac{1}{s}\left(\frac{48}{\pi}\right)^{\frac{1}{3}}\frac{1}{a}\left(1 - \frac{1}{\varepsilon(s)}\right) \quad (S5)$$

$$\varepsilon(s) = \varepsilon_\infty + \frac{\varepsilon_0 - \varepsilon_\infty}{1 + s\tau_D} \quad (S6)$$

$$\varepsilon_0 - \varepsilon_\infty = \frac{A}{\exp\left(\frac{E_a}{kT}\right) - 1} \quad (S7)$$

Figure S7 shows the fitting of the data. In calculation of equation S7 (equation 4 in the main text) for $\varepsilon_0 - \varepsilon_\infty$, $\varepsilon_\infty$ is taken as 4 from experimental data as discussed in section C, the prefactor A is taken as 6.5, and Ea is taken as 20 meV. The solvation cell radius a is determined to be about 10 nm.

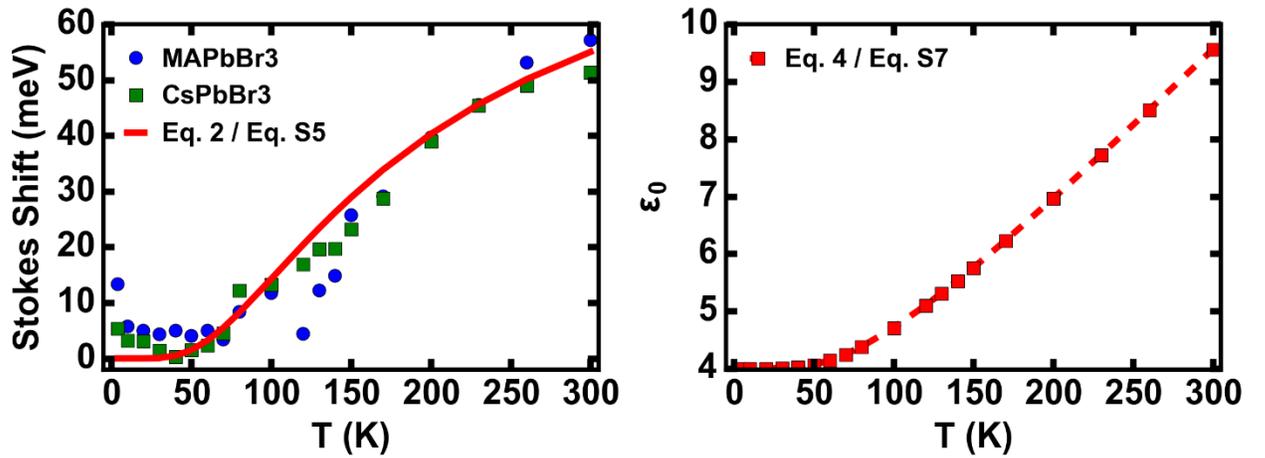

**Fig. S7. Predictions of the point charge solvation model**. Left: Point charge solvation model fitting of the T dependent Stokes shift. Blue dots and green squares denote Stokes shift of MA and Cs observed in experiments respectively. Red curve shows the Stokes shift predicted via dielectric solvation of point charge. Right: Low frequency limit dielectric function $\varepsilon_0$ revealed by fitting the point charge solvation model to T dependent Stokes shift.

*Point dipole solvation model for fitting the T dependent Stokes shift*

The nature of the luminescing band edge species at room temperature has been a topic of ongoing discussion. To account for the stabilization of a possible complex formed by colliding free carriers, we also model the dielectric solvation of a point dipole. The procedure is similar to that of the point charge. For a point dipole, the analytical solution of solvation dynamics in the Laplace domain is expressed in equation S8, where p is the dipole moment of excited species, v is the solvation volume, $\varepsilon(s)$ is the dielectric function, and s = iω. The description of solvent dielectric response function and its temperature dependence follow from the previous case of point charge solvation, expressed in equation S6 and equation S7.

$$\Delta \tilde{E}(s) = \frac{1}{s}\left(\frac{8\pi p^2}{3v}\right)\frac{\varepsilon(s) - 1}{2\varepsilon(s) + 1} \quad (S8)$$

Figure S8 shows the results calculated from the above model to match experimental data. In calculation of equation S7 for $\varepsilon_0 - \varepsilon_\infty$, $\varepsilon_\infty$ is taken as 4 from optical data as discussed in section C, the prefactor A is taken as 6.5, and $E_a$ is taken as 20 meV, identical to the case of point charge solvation. If one takes the e-h separation as 10 nm, then the required solvation volume is about $2.2 \times 10^3$ nm$^3$. The predicted $\varepsilon_0$ shown here is identical to Figure S7.

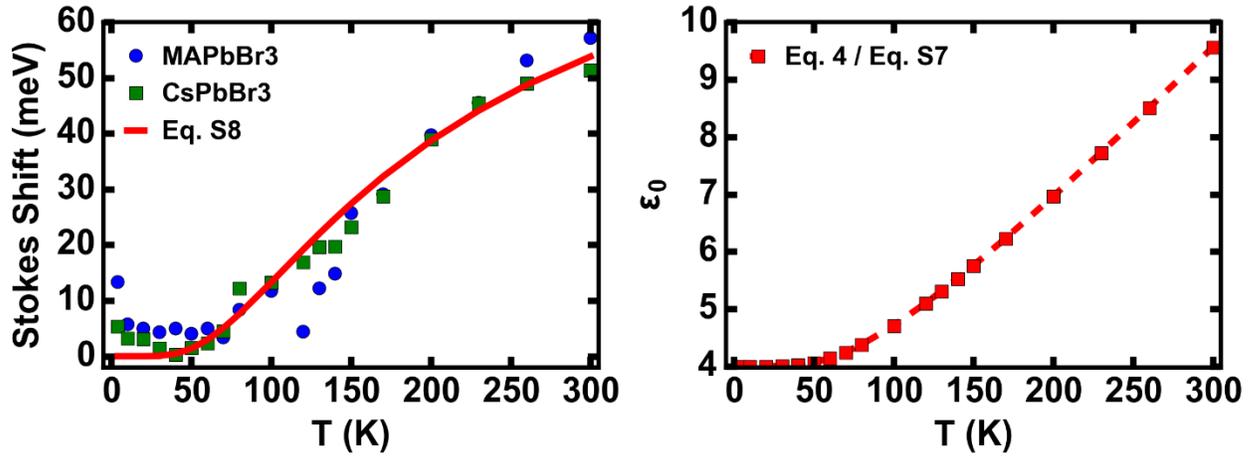

**Fig. S8. Predictions of the point dipole solvation model.** Left: Point dipole solvation model fitting of the T dependent Stokes shift. Blue dots and green squares denote Stokes shift of MA and Cs observed in experiments respectively, identical to Figure S7. Red curve shows the Stokes shift predicted via dielectric solvation of point dipole. The numerical results are very similar to what is shown in Figure S7. Right: Low frequency limit dielectric function $\varepsilon_0$ revealed by fitting the point dipole solvation model to T dependent Stokes shift.

**References**

(1)  Yaffe, O.; Guo, Y.; Tan, L. Z.; Egger, D. A.; Hull, T.; Stoumpos, C. C.; Zheng, F.; Heinz, T. F.; Kronik, L.; Kanatzidis, M. G.; et al. Local Polar Fluctuations in Lead Halide Perovskite Crystals. *Phys. Rev. Lett.* **2017**, *118* (13), 136001.
(2)  Guo, Y.; Yaffe, O.; Paley, D. W.; Beecher, A. N.; Hull, T. D.; Szpak, G.; Owen, J. S.; Brus, L. E.; Pimenta, M. A. Interplay between Organic Cations and Inorganic Framework and Incommensurability in Hybrid Lead-Halide Perovskite CH3NH3PbBr3. *Phys. Rev. Materials* **2017**, *1* (4), 42401.
(3)  Malitson, I. H. Interspecimen Comparison of the Refractive Index of Fused Silica.

*J. Opt. Soc. Am.* **1965**, *55* (10), 1205–1209.
(4) Newville, M.; Stensitzki, T.; Allen, D. B.; Ingargiola, A. LMFIT: Non-Linear Least-Square Minimization and Curve-Fitting for Python, 2014.
(5) Kuzmenko, A. B. Kramers–Kronig Constrained Variational Analysis of Optical Spectra. *Review of Scientific Instruments* **2005**, *76* (8).
(6) Mak, K. F.; He, K.; Lee, C.; Lee, G. H.; Hone, J.; Heinz, T. F.; Shan, J. Tightly Bound Trions in Monolayer MoS2. *Nat Mater* **2013**, *12* (3), 207–211.
(7) Li, Y.; Chernikov, A.; Zhang, X.; Rigosi, A.; Hill, H. M.; van der Zande, A. M.; Chenet, D. A.; Shih, E.-M.; Hone, J.; Heinz, T. F. Measurement of the Optical Dielectric Function of Monolayer Transition-Metal Dichalcogenides: MoS2, MoSe2, WS2, and WSe2. *Phys. Rev. B* **2014**, *90* (20), 205422.
(8) Miyata, K.; Meggiolaro, D.; Trinh, M. T.; Joshi, P. P.; Mosconi, E.; Jones, S. C.; De Angelis, F.; Zhu, X.-Y. Large Polarons in Lead Halide Perovskites. *Science Advances* **2017**, *3* (8).
(9) Tilchin, J.; Dirin, D. N.; Maikov, G. I.; Sashchiuk, A.; Kovalenko, M. V.; Lifshitz, E. Hydrogen-like Wannier-Mott Excitons in Single Crystal of Methylammonium Lead Bromide Perovskite. *ACS Nano* **2016**, *10* (6), 6363–6371.
(10) Abate, J.; Whitt, W. A Unified Framework for Numerically Inverting Laplace Transforms. *INFORMS Journal on Computing* **2006**, *18* (4), 408–421.
(11) Homes, C. C.; Vogt, T.; Shapiro, S. M.; Wakimoto, S.; Ramirez, A. P. Optical Response of High-Dielectric-Constant Perovskite-Related Oxide. *Science* **2001**, *293* (5530), 673–676.
(12) Frohlich, H. *Theory of Dielectrics; Dielectric Constant and Dielectric Loss.*; Clarendon Press: Oxford, 1949.